\documentclass[
    ,final            
  ]
  {aipproc}

\layoutstyle{6x9}

\def\slash#1{#1\!\!\!\!\!/\!\,\,}
\def\Dslash{\slash D}

\newcommand{\be}{\begin{equation}}
\newcommand{\ee}{\nonumber\end{equation}}
\def\beann{\begin{eqnarray*}} \def\eeann{\end{eqnarray*}}

\def\lsim{\raise0.3ex\hbox{$<$\kern-0.75em\raise-1.1ex\hbox{$\sim$}}}
\def\gsim{\raise0.3ex\hbox{$>$\kern-0.75em\raise-1.1ex\hbox{$\sim$}}}

\begin{document}

\title{
{\small \normalfont \vspace*{-2.0cm} \hspace*{10.0cm} CERN-PH-TH/2006-244 \vspace*{1.0cm}} \\
Localization properties of fermions and bosons
}


\author{Philippe de Forcrand}{
address={Institut f\"ur Theoretische Physik, ETH Z\"urich, CH-8093 Z\"urich, Switzerland}
,altaddress={CERN, Physics Department, TH Unit, CH-1211 Geneva 23, Switzerland}
}

\begin{abstract}
The topological structure of the QCD vacuum can be probed by monitoring the
spatial localization of the low-lying Dirac eigenmodes. This approach can be pursued
on the lattice, and unlike the traditional one requires no smoothing of the gauge field.
I review recent lattice studies, attempting to extract a consistent description.
What emerges is a picture of the vacuum as a ``topological sandwich'' of alternating,
infinitely thin $3d$ layers of opposite topological charge,
as originally seen in direct measurements of the topological charge density.
\end{abstract}

\maketitle


``Understanding'' confinement, by identifying the relevant infrared degrees of freedom
of the gauge field, has been a long-standing theoretical goal. It is natural to 
associate this non-perturbative phenomenon with non-perturbative, topological excitations.
The standard list of potentially relevant excitations consists of instantons, Abelian
monopoles and center vortices, with co-dimension 4, 3, and 2 respectively. They each have
received a fluctuating degree of attention over the years. One may hope that a proper
lattice study may unambiguously identify the right excitation. However, in the past it
has been necessary to filter out UV fluctuations of the gauge field in order to reveal
the large-scale structure. This is accomplished by a smoothing/cooling/smearing 
procedure which reduces the action, and inevitably drives the gauge field towards an
action minimum, i.e. a classical instanton solution. Recently, as outlined below,
a different strategy has been followed, which avoids such bias.
The localization properties of low-lying Dirac eigenmodes presumably tell us about
the underlying gauge field excitations, responsible for chiral symmetry breaking and
confinement.

\section{Localization: Anderson and Diakonov-Petrov}

Anderson~\cite{Anderson} considered the Hamiltonian $H = \Delta + V$, where $\Delta$
is a nearest-neighbor hopping operator (a discretized Laplacian) and $V$ a random
potential. This Hamiltonian mimics that of a crystal doped with random impurities.
An eigenmode $\psi(\vec{r})$ can be {\em localized}, meaning that $|\psi(\vec{r})|$ decays
exponentially for large $|\vec{r}|$: the electron cannot hop to infinity, and this mode
does not contribute to the electric conductivity of the material. Otherwise, the 
eigenmode is {\em extended}. Anderson showed that eigenmodes were always localized,
if the disorder in $V$ was large enough, or the energy low enough.
This is intuitively clear: at very high energy, the random potential plays no role
and eigenmodes are extended, plane-wave-like; but at low energy, sufficient randomness in
the potential may forbid hopping to any of the neighboring sites. Thus, the spectrum
looks generically as in Fig.~1, with a {\em mobility edge} $\lambda_c$  separating the
localized from the extended regime.
$\lambda_c$ may lie above or below the ground-state energy depending on the disorder,
allowing for a so-called quantum transition at zero temperature.

\begin{figure}
  \includegraphics[width=0.8\textwidth]{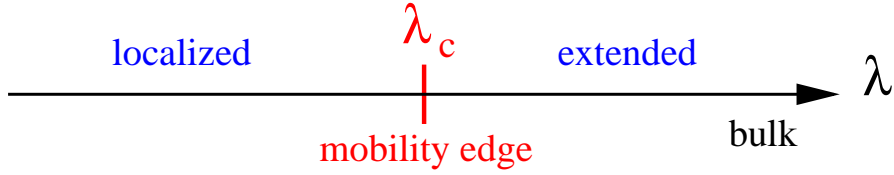}
  \caption{Localization properties of boson eigenstates in a random medium.}
\end{figure}

Diakonov and Petrov~\cite{Diakonov} proposed an analogous explanation for the QCD
chiral transition at temperature $T_c$. Since 
$\langle \bar\psi \psi \rangle = -\pi~ \rho(0)$~\cite{B-C}, 
the spectral properties of the Dirac operator, described by the density $\rho(\lambda)$,
must change with temperature. Recalling that an instanton supports a chiral zero mode $\psi^I$
(for each fermion flavor), one considers the Dirac spectrum for a linear superposition of
Instantons and Antiinstantons. The exact zero modes are now displaced. The infrared
spectrum results from diagonalizing the effective Dirac matrix 
$\left( \begin{array}{cc}0 & T_{IA} \\ T_{IA}^\dagger & 0 \end{array} \right)$
of overlap elements $T_{ij} = \langle \psi^I_i | \psi^A_j \rangle$ between individual Instantons
and Antiinstantons zeromodes. At $T=0$, this overlap decreases as $\frac{1}{|r_{ij}|^3}$, and
gives rise to extended modes having support on all instantons and antiinstantons, and
an essentially uniform spectrum near zero. Such extended modes can be observed on
the lattice (see Fig.~2).
As the temperature increases, the density of instantons decreases (their action
$8\pi^2/g^2(T)$ increases) and $T_{ij}$ now decays exponentially $\sim \exp(-\pi r_{ij} T)$
in spatial directions. Both factors may trigger a transition to localization, which
suppresses near-zero eigenvalues and restores chiral symmetry.

\begin{figure}
  \includegraphics[width=.7\textwidth]{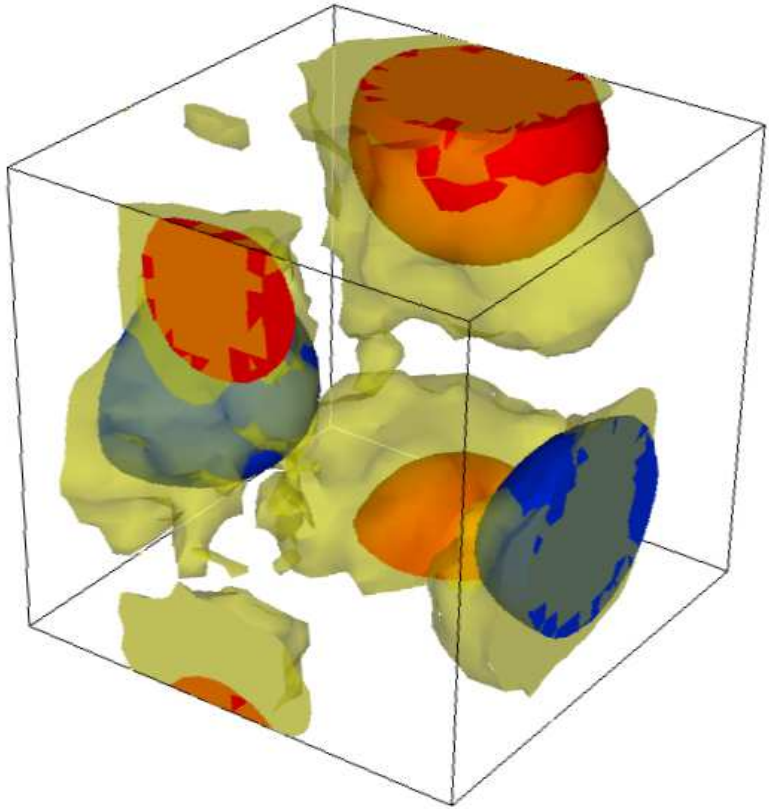}
  \caption{Surface (yellow) of equal magnitude of lowest-lying Dirac eigenmode 
in the presence of an Instanton-Antiinstanton pair (red and blue spheroids), 
from \cite{old}. The eigenmode is {\em extended} over both objects.
The gauge field was cooled 
to identify the $I-A$ pair, but {\em not} to obtain the Dirac eigenmode.}
\end{figure}

As in Anderson's bosonic case, this is a disorder-driven transition. But Dirac
eigenvalues come in pure imaginary pairs $\pm i \lambda$, and the focus here
is on the center of the spectrum $\lambda=0$. Correspondingly, the transition can be
modeled by {\em chiral} random matrices~\cite{G-G}.

It is important to note that the Diakonov-Petrov scenario does not require instantons,
but only chiral zero modes. It turns out that other topological defects - domain-walls,
monopoles, vortices - also support chiral zero modes. Under the working assumption that
extended modes have support on the union of topological defects, this opens the
possibility to determine the topological vacuum structure from the spatial distribution
of low-lying eigenmodes. This approach is gauge-invariant and requires no smoothing/cooling
of the gauge field.

\section{Lattice studies}

To characterize how localized or extended a mode $\psi(x)$ is, one uses the {\em inverse
participation ratio} $IPR \equiv V \frac{\sum_x |\psi(x)|^4}{(\sum_x |\psi(x)|^2)^2}$.
This ratio of moments is equal to 1 if $\psi(x)=\delta(x_0)$ is completely localized,
and to $V$ if $\psi(x)=const.~\forall x$ is completely delocalized. If $\psi(x)=1$ on
a fraction $f$ of the sites, 0 elsewhere, then $IPR=1/f$, which justifies its name.

Of course one should consider the continuum limit $a\to 0$ of the lattice study.
Note that the fraction $f$ of occupied sites, hence the IPR, remains constant as $a\to 0$ for 
any type of macroscopic object. Conversely, if the object is ``thin'' and lives on a
submanifold of dimension $d$ and volume ${\cal V}_d$, then 
$f = \frac{{\cal V}_d / a^d}{V/a^4} \sim a^{4-d}$, 
and $IPR \sim a^{d-4}$, which diverges as $a\to 0$ (see Fig.~3).
$d=0, 1, 2$ then characterizes ``thin'' instantons (point-like), monopoles (line-like)
and vortices (surface-like).

\begin{figure}
  \includegraphics[width=.23\textwidth]{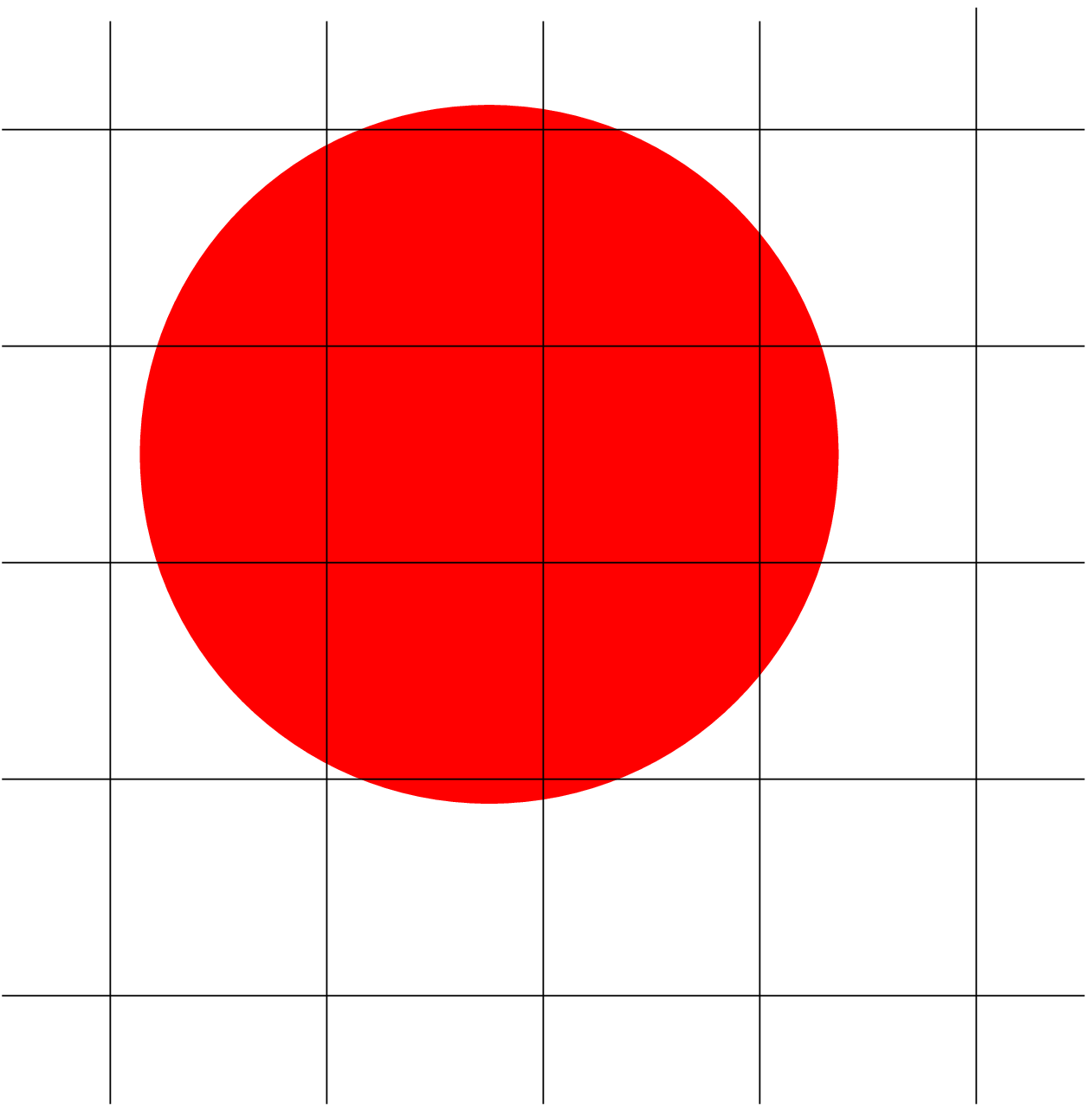}
  \includegraphics[width=.23\textwidth]{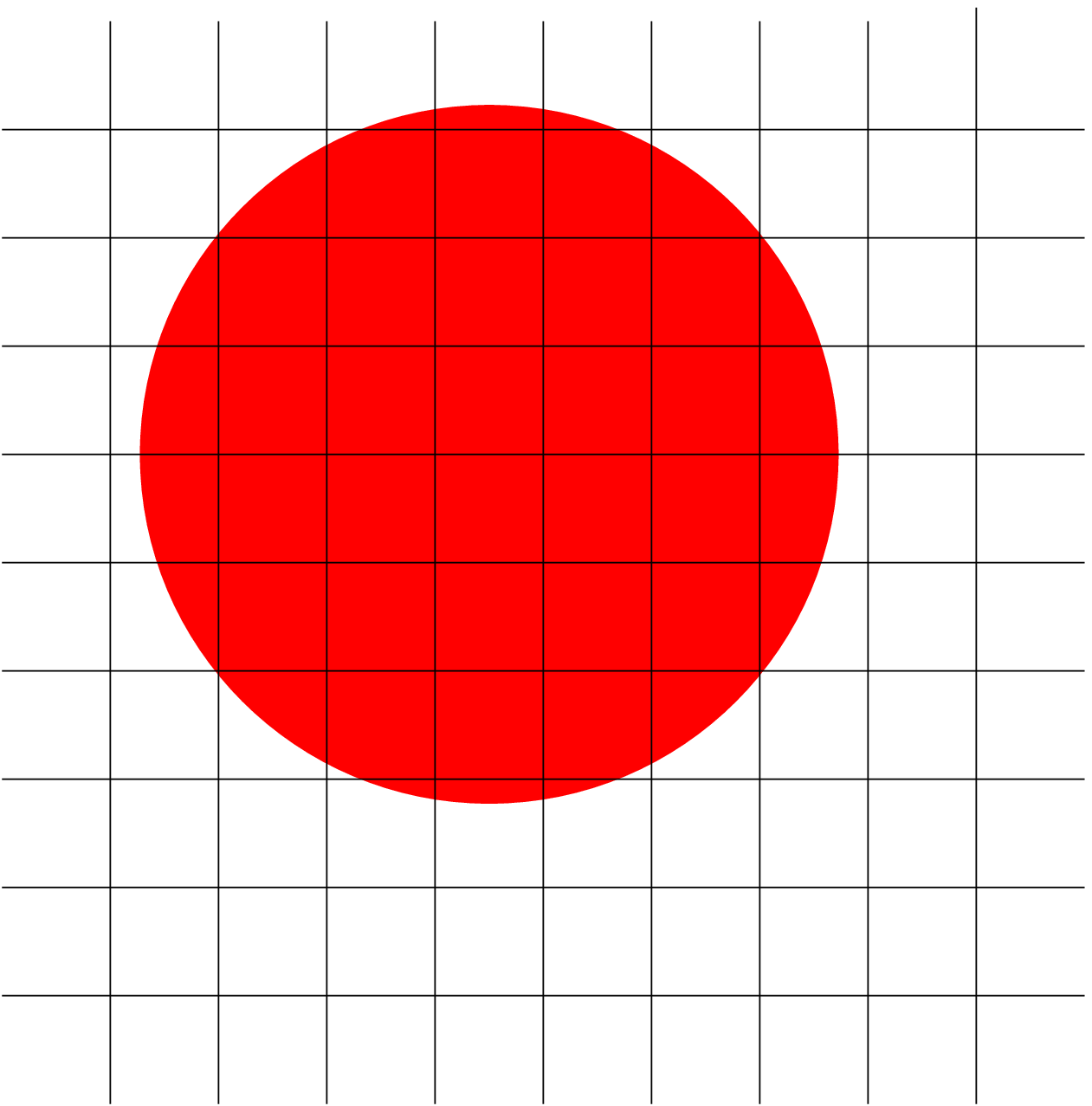}
  \includegraphics[width=.23\textwidth]{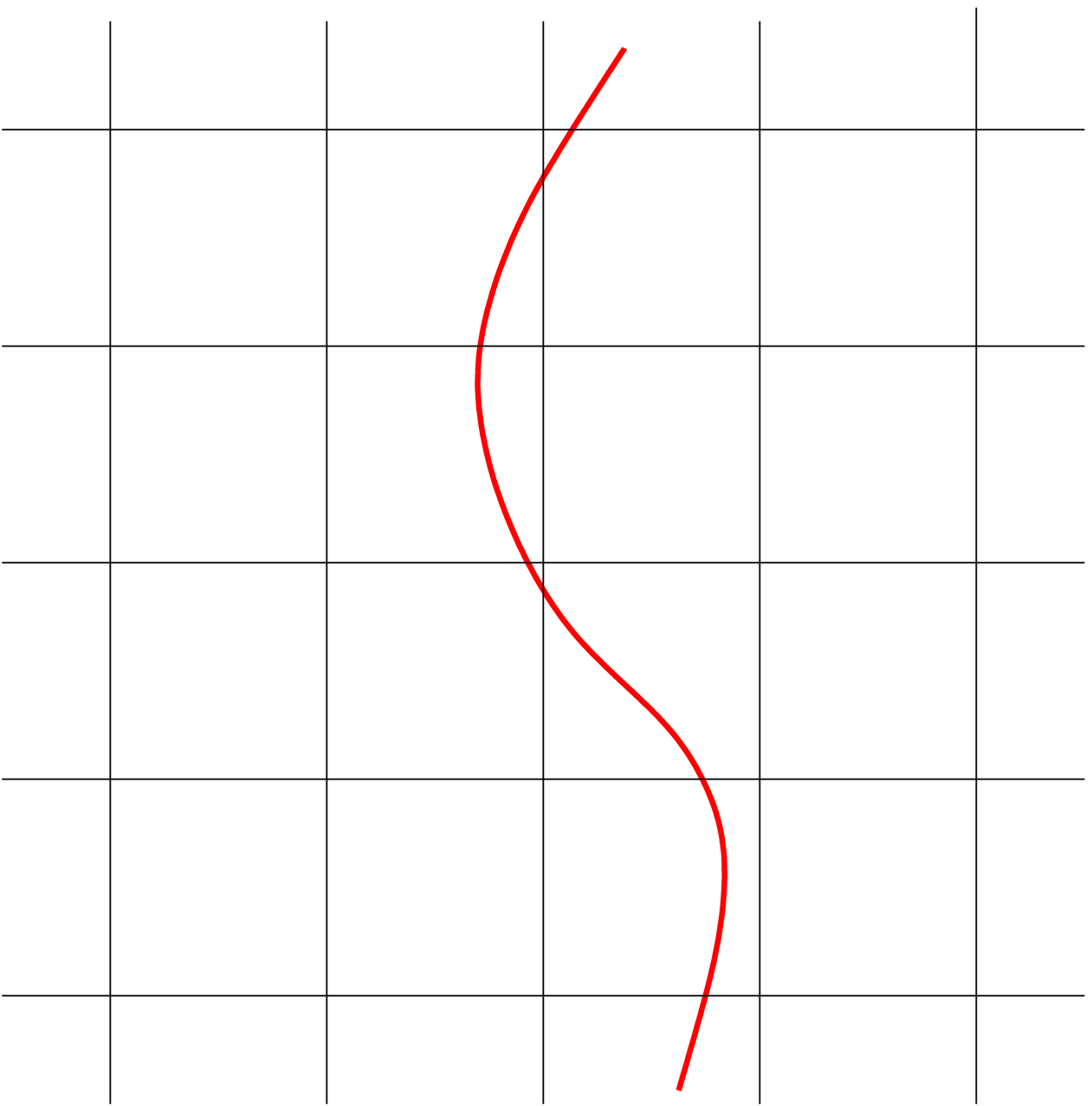}
  \includegraphics[width=.23\textwidth]{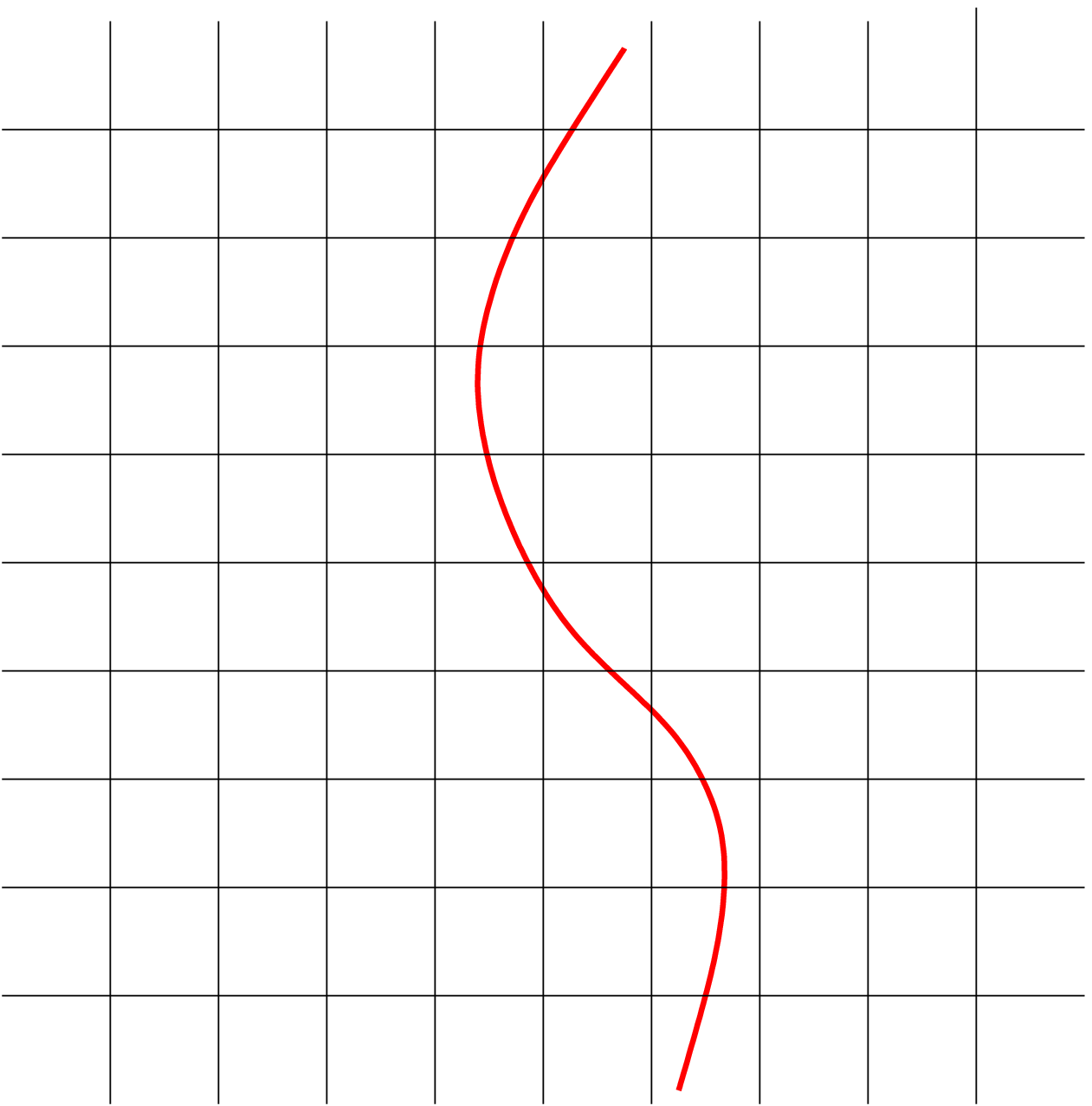}
  \caption{The fraction $f$ of occupied lattice sites, hence the $IPR$, remains 
unchanged as $a\to 0$ in the presence of {\em any} thick object ({\em left}). 
It goes to zero as $a^{d-4}$ for a thin object of dimension $d$ ({\em right}).
\vspace*{-0.3cm}}
\end{figure}

$\bullet$ The first study, by the MILC collaboration~\cite{MILC} for $SU(3)$, indicated
$d\sim 3$: the vacuum, it seems, is made of infinitely thin domain-walls! (see Fig.~4, left).

\begin{figure}
  \includegraphics[width=0.45\textwidth]{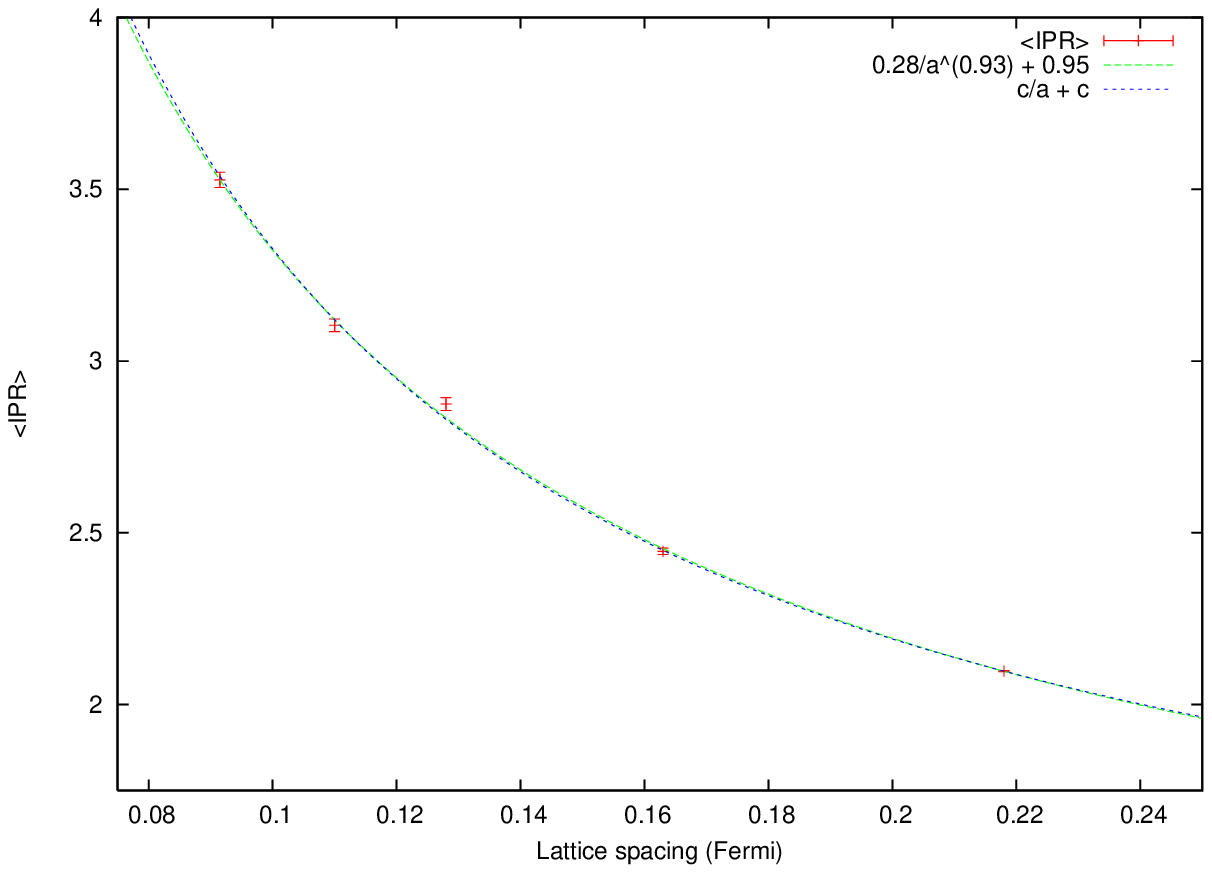}
\put(-100,100){\small $d = 3.07 \pm 0.15$}
  \includegraphics[width=0.45\textwidth]{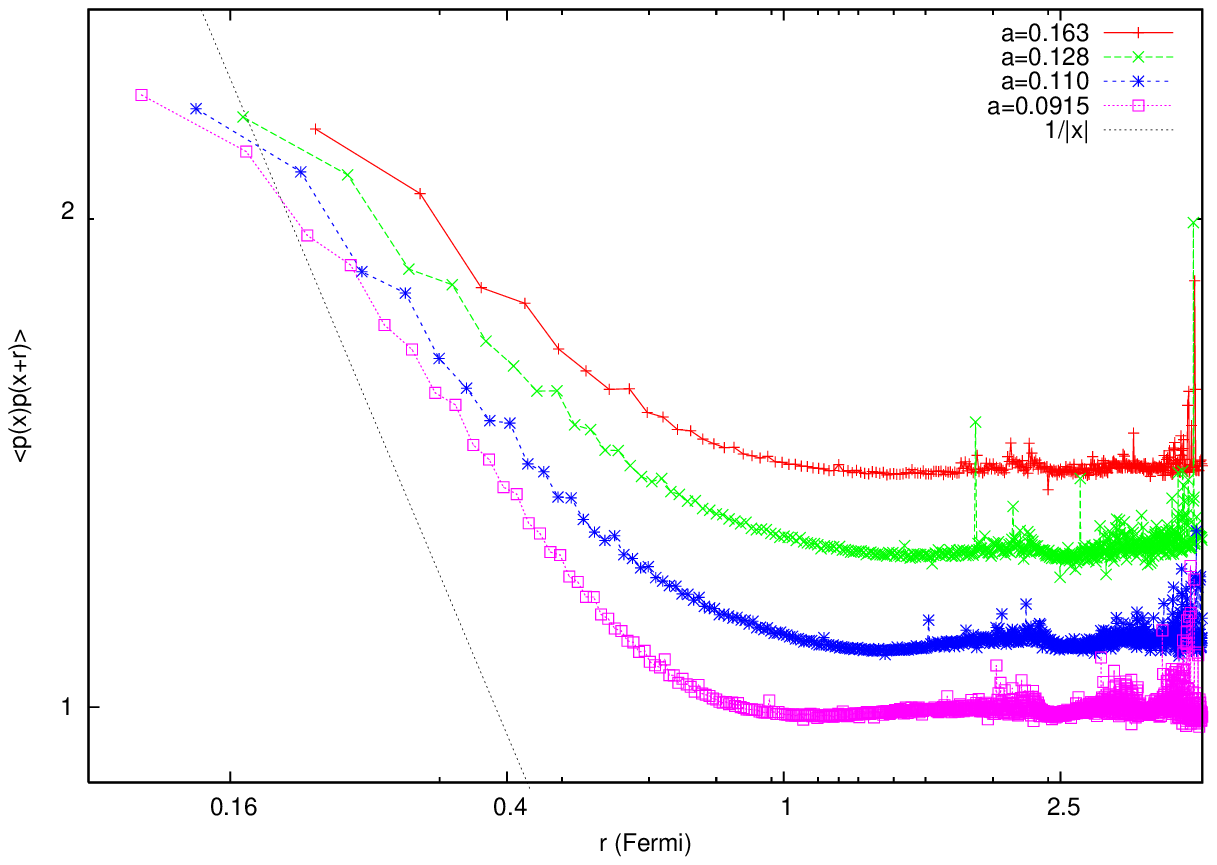}
  \caption{({\em left}) Most recent $IPR$ data from MILC Collaboration \cite{MILC} versus
lattice spacing, for $SU(3)$ with Asqtad Dirac operator and Symanzik gauge action. The dimension 
of the supporting manifold is about 3.
({\em right}) Correlator $\langle |\psi(0)| |\psi(\vec{x})| \rangle$ of magnitude of low-lying Dirac 
eigenmode, for different lattice spacings.
The straight line shows what would happen if $|\Psi|=1$ on a $3d$ fractal, 0 elsewhere.}
\end{figure}

$\bullet$ This surprising result was quickly checked in \cite{Zakharov}. There, the IPR diverges 
even faster, favoring point-like instantons (see Fig.~5). That study used the overlap discretization
of the Dirac operator, with exact chiral properties, but used the
Wilson action for the gauge group $SU(2)$. Unfortunately, it is known since \cite{Teper}
that with this choice of gauge action, the action of size $a$ defects (``dislocations'')
is insufficient to compensate for their entropy $\sim \log a^{-4}$, and they become
dense as $a\to 0$. The observations of \cite{Zakharov} may then be entirely caused by
lattice artifacts.

\begin{figure}
  \includegraphics[width=.45\textwidth]{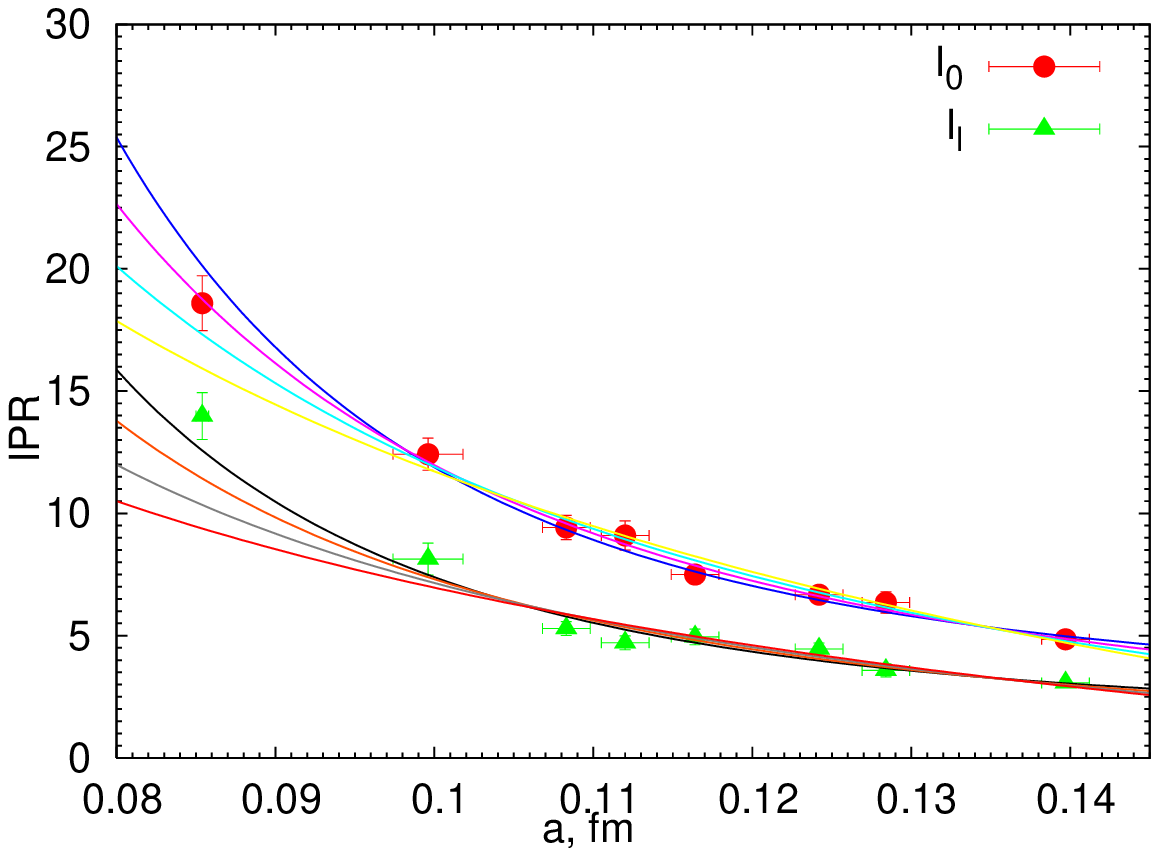}
  \includegraphics[width=.45\textwidth]{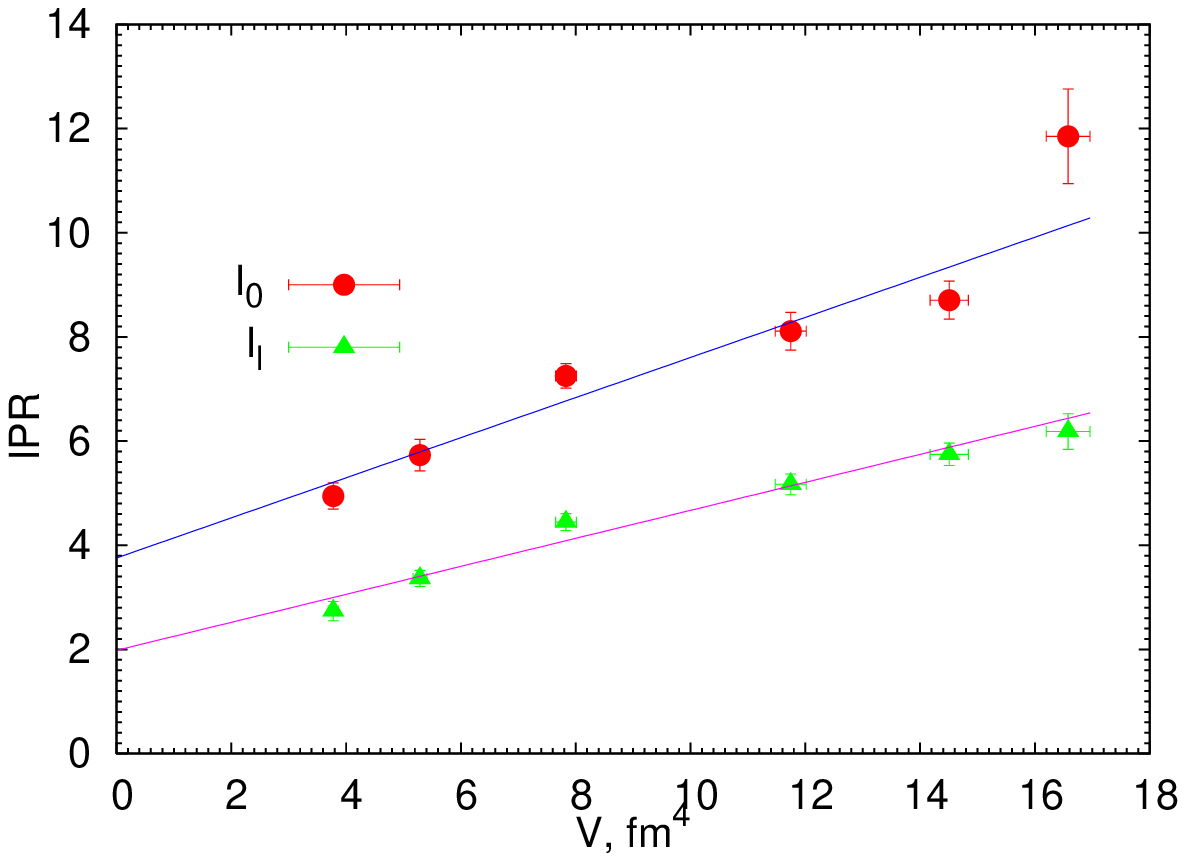}
\put(-67,68){\tiny 0-modes}
\put(-74,45){\tiny near-0 modes}
\put(-305,58){\tiny 0-modes}
\put(-310,32){\tiny near-0 modes}
\put(-243,85){\tiny $d=0$}
\put(-234,78){\tiny $1$}
\put(-234,71){\tiny $2$}
\put(-234,64){\tiny $3$}
  \caption{$IPR$ versus lattice spacing ({\em left}) and volume ({\em right}), for $SU(2)$
with overlap Dirac operator and Wilson gauge action \cite{Zakharov}.
\vspace*{-0.2cm}}
\end{figure}

$\bullet$ A third study, also with the overlap Dirac operator, but for $SU(3)$ and
L\"uscher-Weisz gauge action, shows results (Fig.~6, left) roughly consistent with \cite{MILC}.
The IPR for the low-lying modes grows more or less as $1/a$, consistent again with thin domain-wall
structures (Fig.~6, right).

\begin{figure}
  \includegraphics[width=.48\textwidth,height=0.30\textwidth]{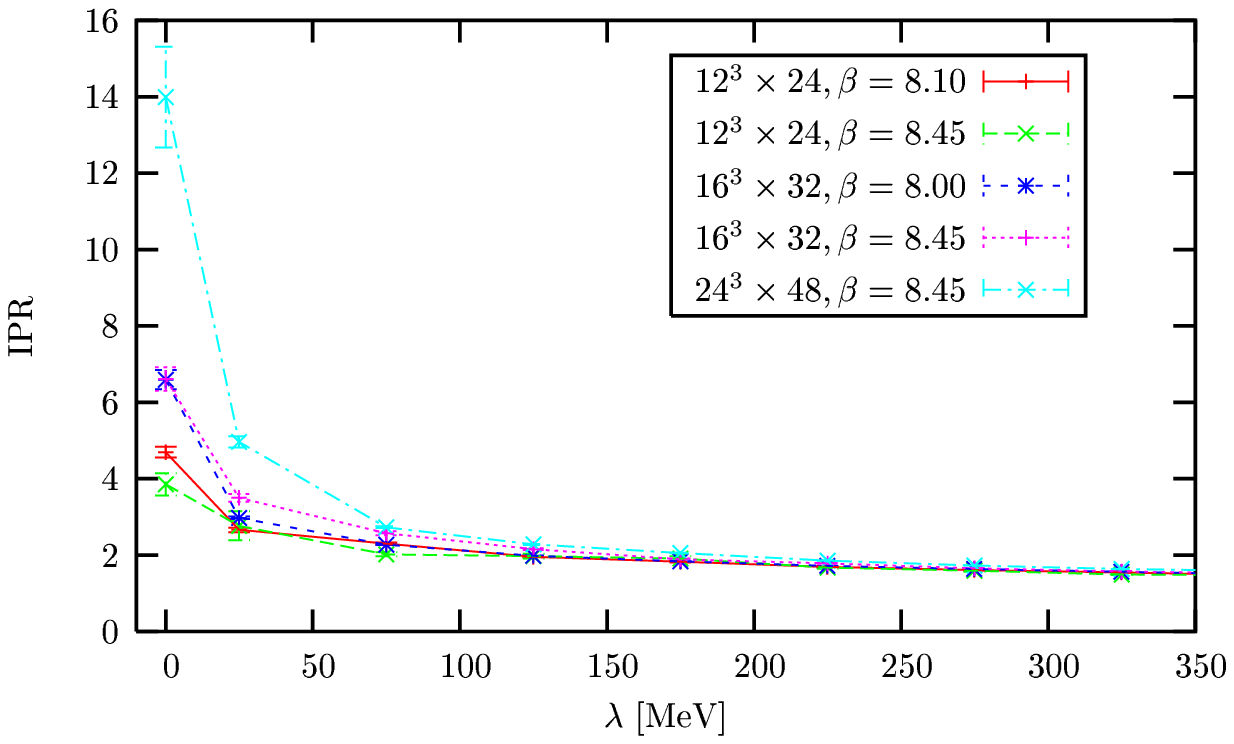}
  \includegraphics[width=.48\textwidth,height=0.31\textwidth]{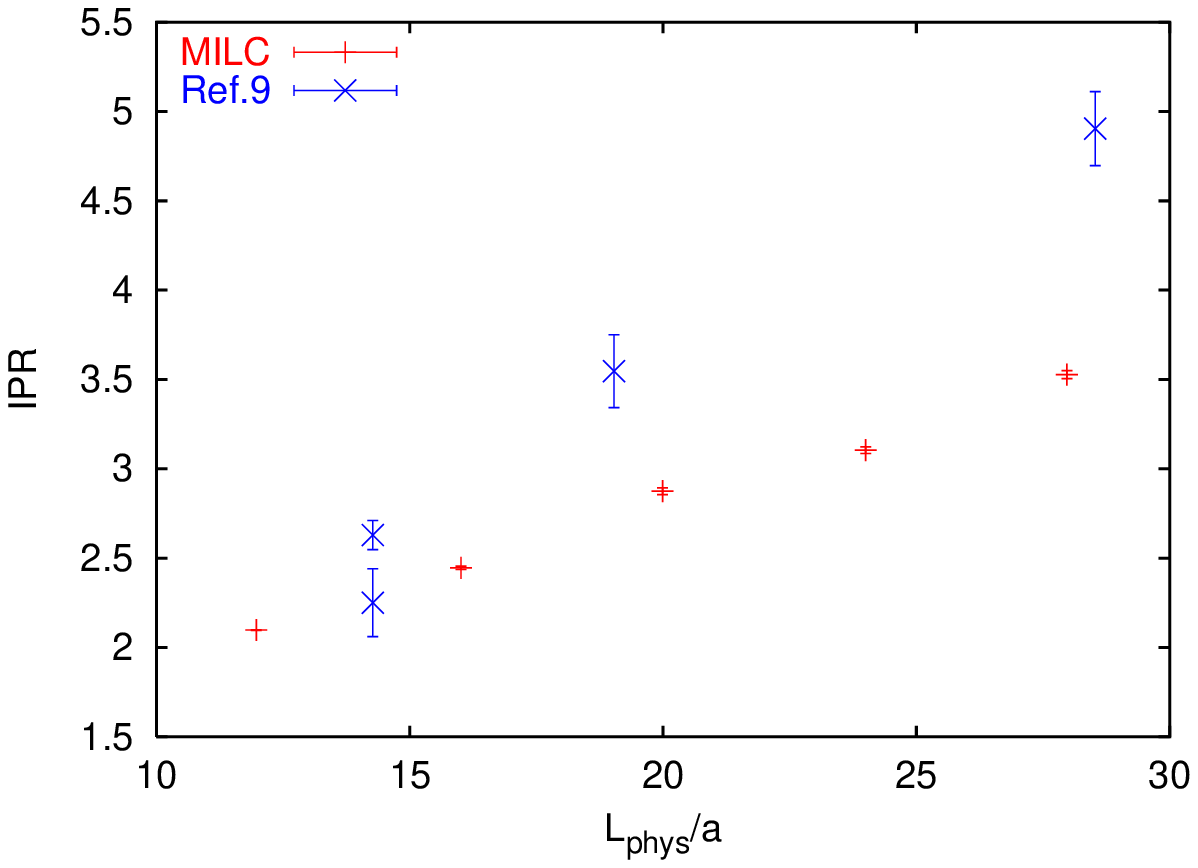}
  \caption{({\em left}) $IPR$ versus eigenvalue and lattice spacing,
for $SU(3)$ with overlap Dirac operator and L\"uscher-Weisz gauge action \cite{Schierholz}.
({\em right}) For the lowest non-zero modes, the $IPR$ grows $\propto 1/a$ as for MILC.}
\end{figure}

\begin{figure}
  \includegraphics[width=0.5\textwidth,height=0.25\textwidth]{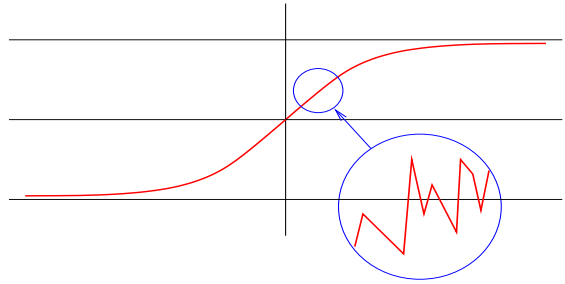}
  \caption{A classical kink is smooth. The actual kink looks like a random walk
at short distances.}
\end{figure}

It should not have come as a surprise that the IPR diverges. Even if classical lumps of size
$1/\Lambda_{QCD}$ are relevant to the QCD vacuum structure, these lumps do not look smooth
at short distance due to quantum fluctuations (see Fig.~7).
Besides, this result can be related to a curious property of the topological
charge density operator $\langle q(0) q(\vec{x}) \rangle$~\cite{Seiler}. 
On one hand, this correlator is negative
for any $\vec{x}\neq 0$. This can be seen from reflection-positivity, or simply by realizing
that $q(\vec{x}) \sim \vec{E} \cdot \vec{B}$ acquires an extra factor $i$ in Euclidean space.
On the other hand, $\langle \int d^4x~ q(0) q(\vec{x}) \rangle = \chi_{\rm top}$ is finite and about
$(190 {\rm MeV})^4$. Hence, a contact term $\delta(\vec{x})$ must compensate the negative $\vec{x}\neq 0$
integral. Moreover, the canonical dimension of $q(\vec{x})$ is 4, so that one expects
$\langle \int_{\vec{x}\neq 0} d^4x~ q(0) q(\vec{x}) \rangle$ to diverge. This divergence is exactly
compensated by the contact term to leave the finite piece $\chi_{\rm top}$.

\begin{figure}
  \includegraphics[width=.45\textwidth,height=0.345\textwidth]{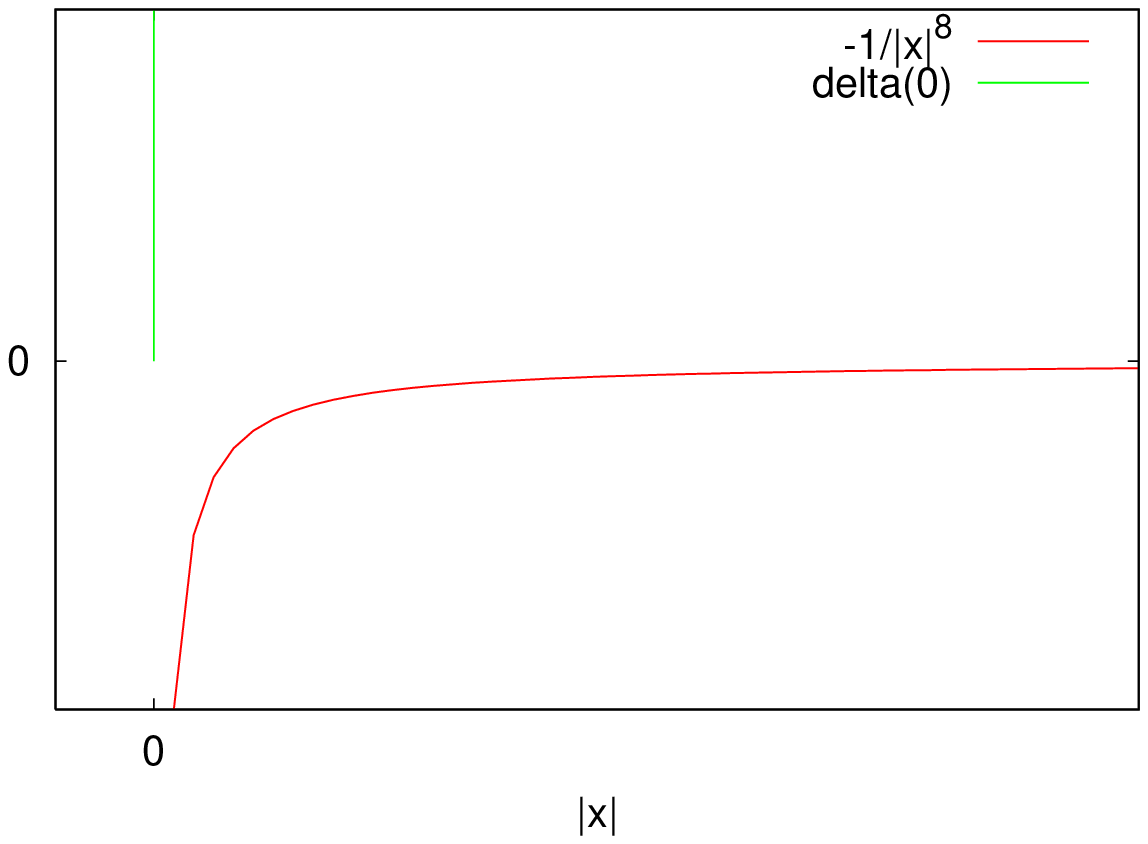}
  \includegraphics[width=.50\textwidth,height=0.330\textwidth]{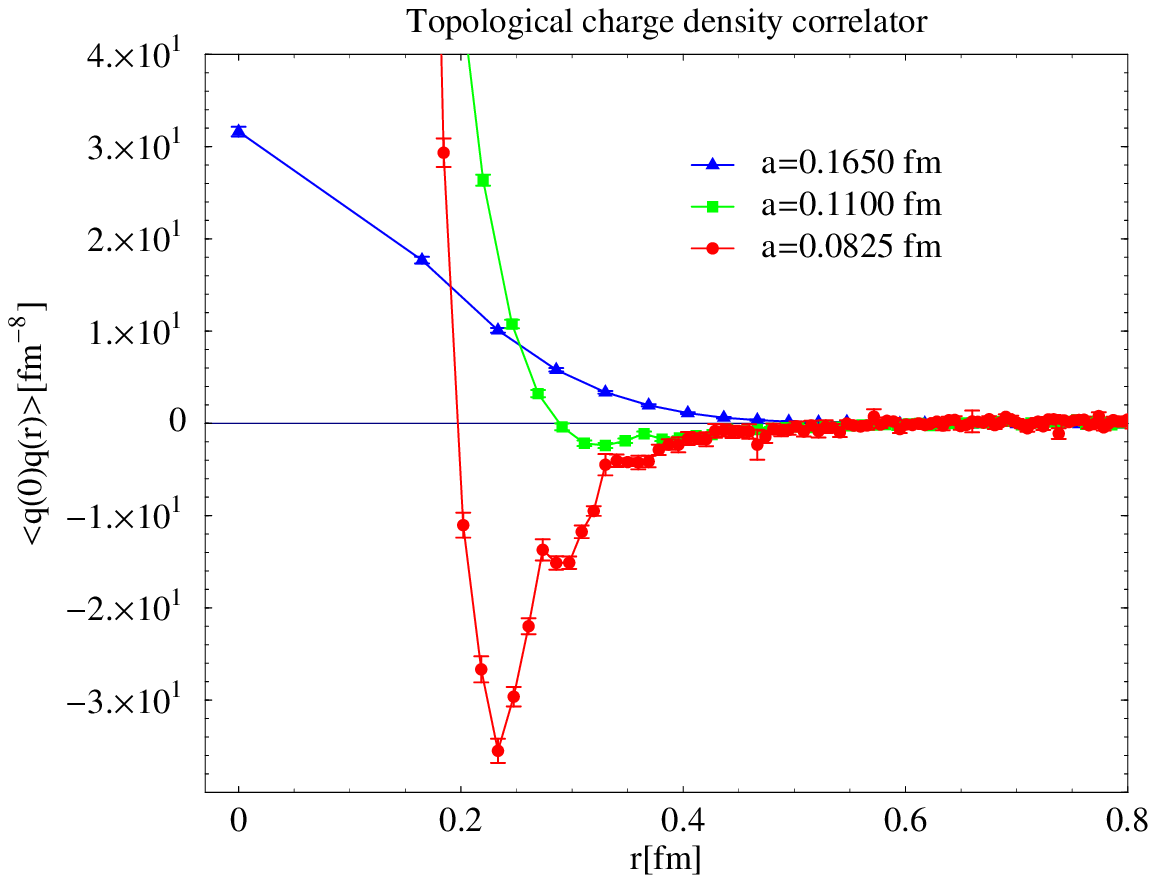}
\put(-105,45){\small $q(x) = {\rm Tr_{color,Dirac}} \gamma_5 \Dslash$}
  \caption{Correlator of the topological charge density: ({\em left}) as expected in the 
continuum and ({\em right}) as measured on the lattice \cite{KFLiu}. The negative
part and divergent contact term both become visible as $a\to 0$.}
\end{figure}

This behaviour has recently been observed on the lattice: compare Fig.~8 left and right.
One has then to understand how the vacuum becomes ``topologically antiferromagnetic'', with 
$\langle q(0) q(\vec{x}) \rangle$ going to $-\infty$ as $|\vec{x}| \to 0$.
Before the studies of Dirac eigenmodes reviewed here,
the Kentucky group already claimed to see space-filling $3d$ structures of transverse size ${\cal O}(a)$
and of opposite topological charges \cite{KFLiu}.
Analogous structures have now been seen in Ref.~\cite{Schierholz}, and also in the $CP^3$ model~\cite{Thacker}
(see Fig.~9). 

\begin{figure}
  \includegraphics[width=.43\textwidth]{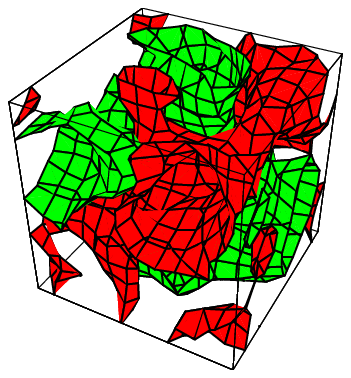}
  \includegraphics[width=.48\textwidth]{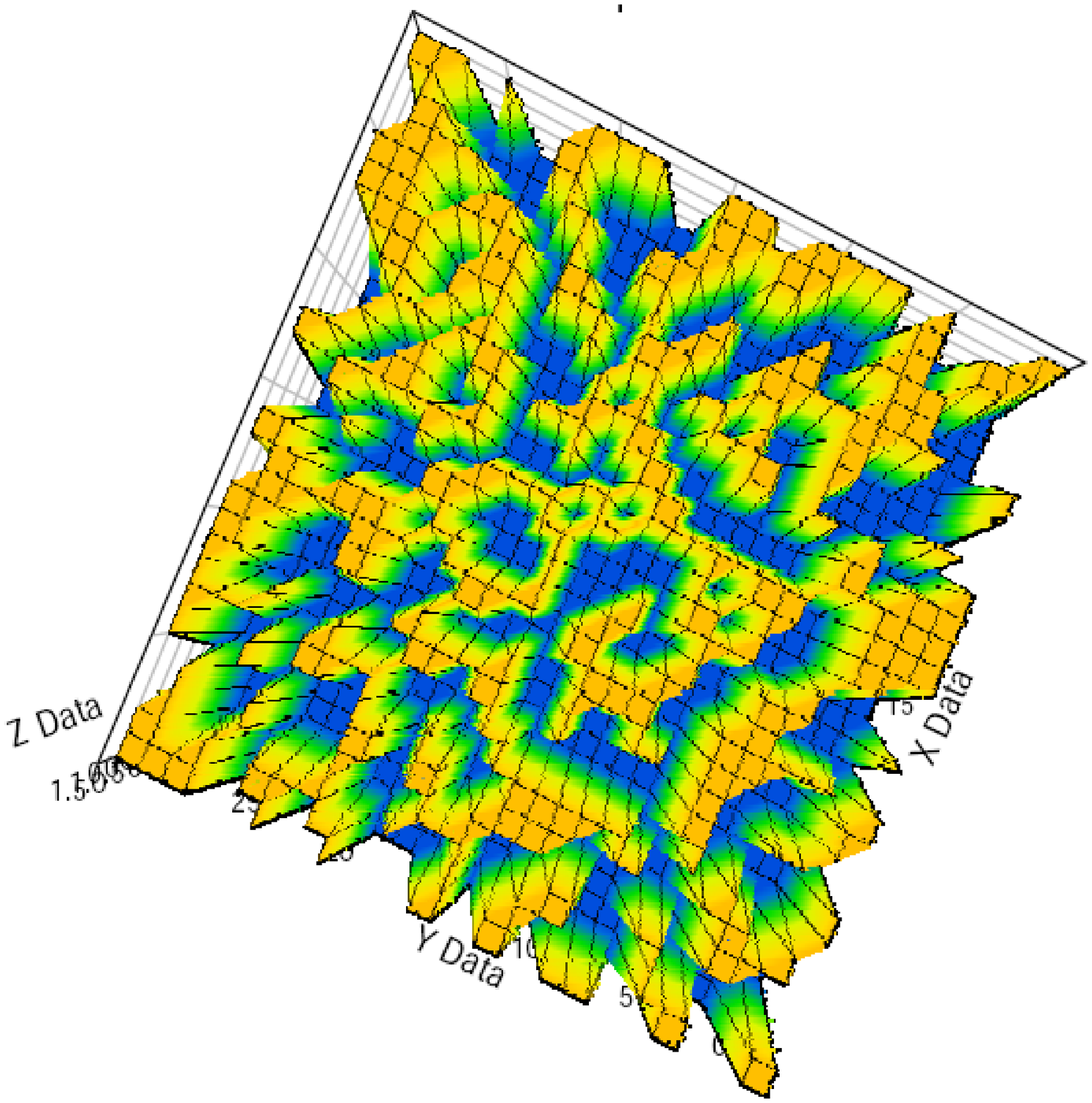}
  \caption{Topological charge density: ({\em left}) in QCD \cite{Schierholz}
and ({\em right}) in the $(1+1)d$ $CP^3$ model \cite{Thacker}.
\vspace*{-0.3cm}}
\end{figure}

As a final check, the MILC collaboration has measured the correlator 
$\langle |\psi(0)| |\psi(\vec{x})| \rangle$, where $\psi$ is a low-lying Dirac eigenmode.
If $|\psi(\vec{x})|=1$ on a $3d$ fractal and 0 elsewhere, this correlator would decay as $1/|\vec{x}|$.
Measurements~\cite{MILC} are not too different (Fig.~4, right), until $|\vec{x}|$ reaches distances
${\cal O}(1/\Lambda_{QCD})$.

\begin{figure}[b!]
  \includegraphics[width=.45\textwidth]{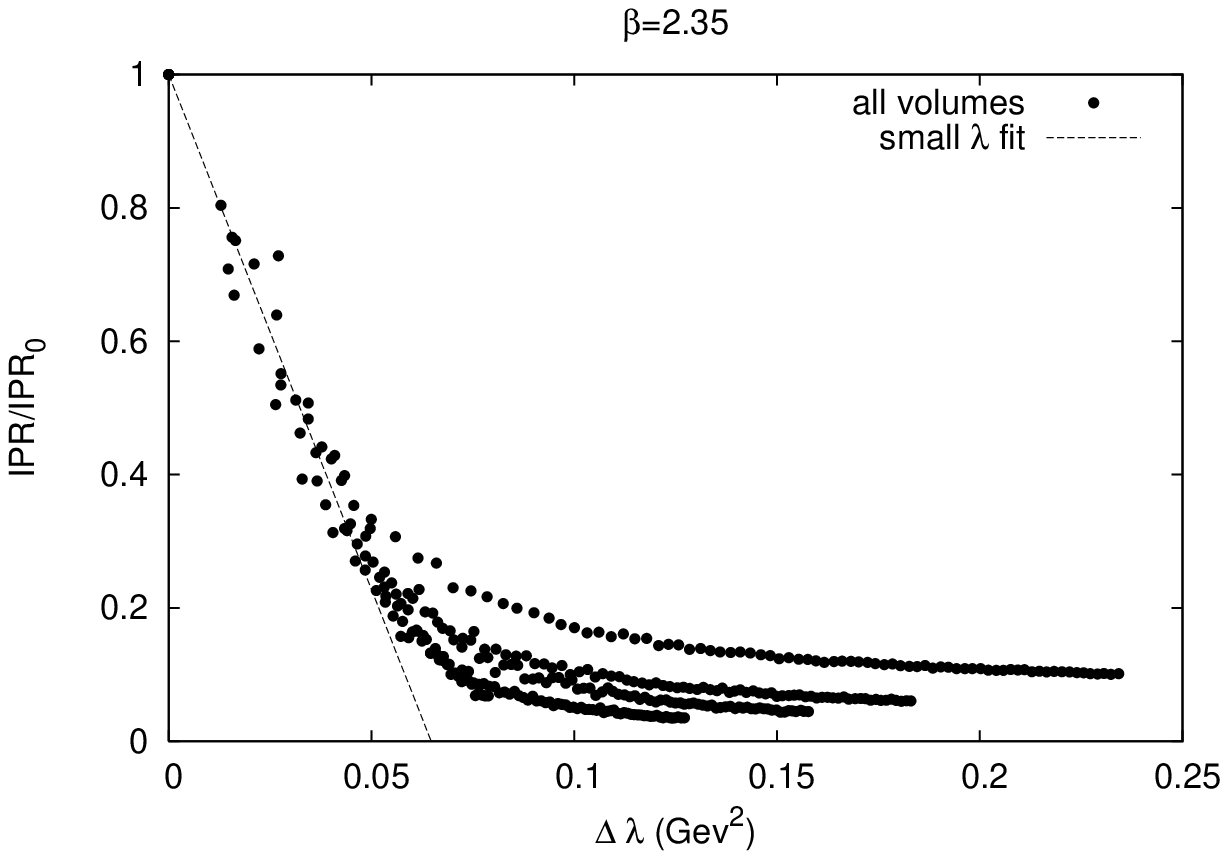}
  \includegraphics[width=.45\textwidth]{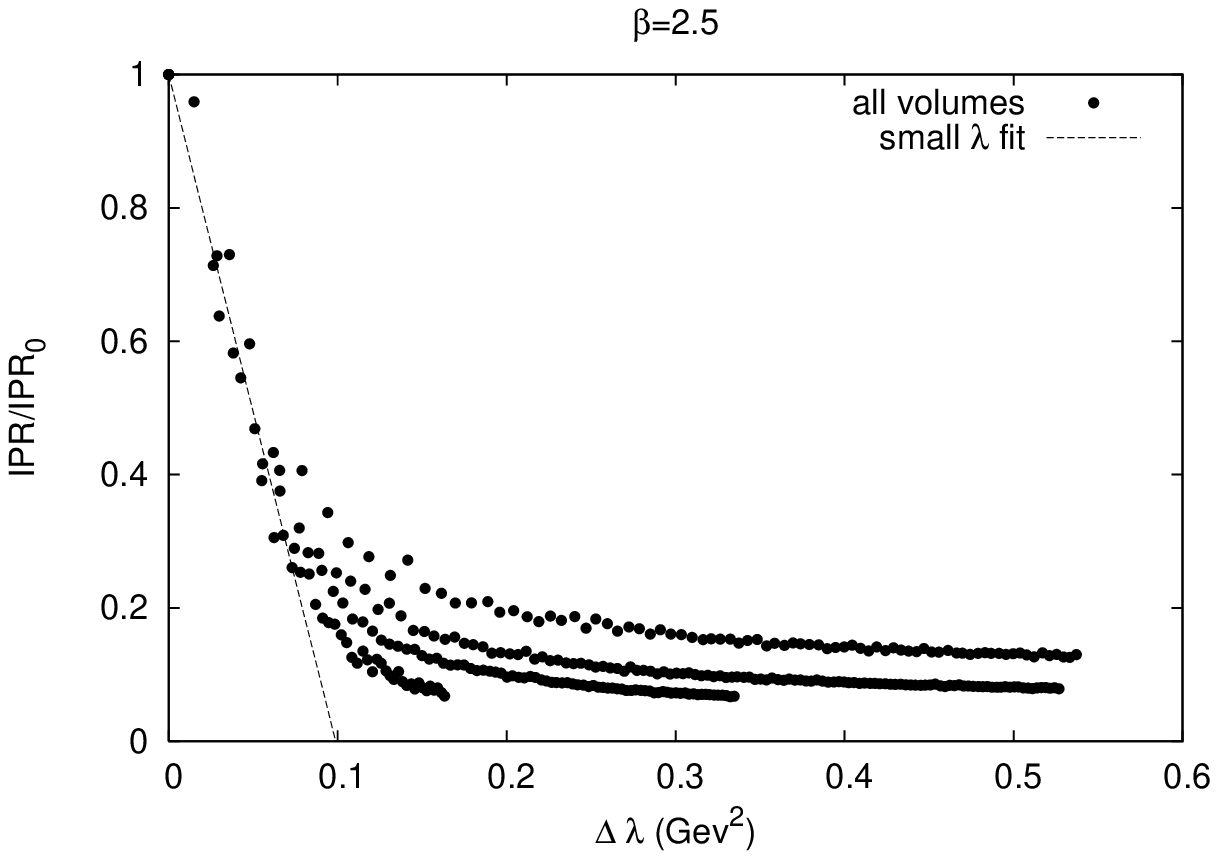}
\put(-270,75){\small large $a$}
\put(-80,75){\small small $a$}
\put(-350,20){\tiny localized}
\put(-290,35){\tiny extended}
\put(-160,20){\tiny localized}
\put(-85,35){\tiny extended}
\put(-139,18){\bf $\downarrow$}
\put(-318,18){\bf $\downarrow$}
  \caption{The mobility edge (arrow) for an adjoint Higgs appears to go to $\infty$ 
as $a\to 0$ \cite{bosons}, for $SU(2)$ with Wilson gauge action.
\vspace*{-1.0cm}}
\end{figure}

$\bullet$ Most recently, this kind of lattice study has been extended to bosonic fields and
eigenmodes of the covariant Laplacian in various representations~\cite{bosons}, with yet more surprises.
In the adjoint representation, the mobility edge is found to rise unexpectedly as $1/a$ (see Fig.~10).
If true, all finite-energy modes in the continuum would be localized, contradicting perturbation
theory. One wonders if this strange result also can be blamed on the dense dislocations caused by the
choice of Wilson action for gauge group $SU(2)$.

\section{Conclusion}

The picture of the QCD vacuum which emerges (Figs.~4, 6, 8, 9) is that of a ``topological sandwich'', with alternating, 
infinitely thin
$3d$ layers of opposite topological charge density. While bizarre, this picture is not forbidden,
and is supported by some theoretical arguments~\cite{Seiler}. But it is far from the usual picture
of a dilute gas of classical excitations. One should stress that these two descriptions are
not mutually exclusive, as sketched Fig.~11: one applies at UV scales, the other may apply at scales
${\cal O}(1/\Lambda_{QCD})$. The structure of the QCD vacuum depends on the scale considered.

\begin{figure}
  \includegraphics[width=0.7\textwidth,height=0.30\textwidth]{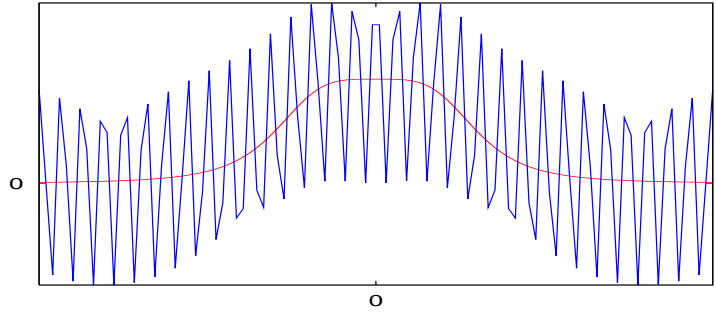}
  \caption{The ``topological sandwich'' structure observed at short distance (blue) is not
inconsistent with another structure (red) at scale $1/\Lambda_{QCD}$.}
\end{figure}


\begin{thebibliography}{99}

\bibitem{Anderson}
  P.~W.~Anderson,
  Phys.\ Rev.\  {\bf 109} (1958) 1492.

\bibitem{Diakonov}
  D.~Diakonov and V.~Y.~Petrov,
  Nucl.\ Phys.\ B {\bf 272} (1986) 457;
  D.~Diakonov,
  arXiv:hep-ph/9602375.

\bibitem{B-C}
  T.~Banks and A.~Casher,
  Nucl.\ Phys.\ B {\bf 169} (1980) 103.

\bibitem{old}
  P.~de Forcrand, M.~Garcia Perez, J.~E.~Hetrick, E.~Laermann, J.~F.~Lagae and I.~O.~Stamatescu,
  Nucl.\ Phys.\ Proc.\ Suppl.\  {\bf 73} (1999) 578
  [arXiv:hep-lat/9810033].

\bibitem{G-G}
  A.~M.~Garcia-Garcia and J.~C.~Osborn,
  Phys.\ Rev.\ Lett.\  {\bf 93} (2004) 132002
  [arXiv:hep-th/0312146].

\bibitem{MILC}
  C.~Aubin {\it et al.}  [MILC Collaboration],
  Nucl.\ Phys.\ Proc.\ Suppl.\  {\bf 140} (2005) 626
  [arXiv:hep-lat/0410024];
  C.~Bernard {\it et al.},
  PoS {\bf LAT2005} (2006) 299
  [arXiv:hep-lat/0510025].


\bibitem{Zakharov}
  F.~V.~Gubarev, S.~M.~Morozov, M.~I.~Polikarpov and V.~I.~Zakharov,
  arXiv:hep-lat/0505016;
  M.~I.~Polikarpov, F.~V.~Gubarev, S.~M.~Morozov and V.~I.~Zakharov,
  PoS {\bf LAT2005} (2006) 143
  [arXiv:hep-lat/0510098].
~ See 
  V.~I.~Zakharov,
  arXiv:hep-ph/0602141 
for an alternative review.

\bibitem{Teper}
  D.~J.~R.~Pugh and M.~Teper,
  Phys.\ Lett.\ B {\bf 224} (1989) 159.


\bibitem{Schierholz}
  Y.~Koma, E.~M.~Ilgenfritz, K.~Koller, G.~Schierholz, T.~Streuer and V.~Weinberg,
  PoS {\bf LAT2005} (2006) 300
  [arXiv:hep-lat/0509164];
  V.~Weinberg, E.~M.~Ilgenfritz, K.~Koller, Y.~Koma, G.~Schierholz and T.~Streuer,
  arXiv:hep-lat/0610087.

\bibitem{Seiler}
  E.~Seiler and I.~O.~Stamatescu,
MPI-PAE/PTh 10/87;
  E.~Vicari,
  Nucl.\ Phys.\ B {\bf 554} (1999) 301
  [arXiv:hep-lat/9901008];
  E.~Seiler,
  Phys.\ Lett.\ B {\bf 525} (2002) 355
  [arXiv:hep-th/0111125].

\bibitem{KFLiu}
  I.~Horvath {\it et al.},
  Phys.\ Rev.\ D {\bf 68} (2003) 114505
  [arXiv:hep-lat/0302009];
  Phys.\ Lett.\ B {\bf 617} (2005) 49
  [arXiv:hep-lat/0504005].

\bibitem{Thacker}
  S.~Ahmad, J.~T.~Lenaghan and H.~B.~Thacker,
  Phys.\ Rev.\ D {\bf 72} (2005) 114511
  [arXiv:hep-lat/0509066].

\bibitem{bosons}
  J.~Greensite, S.~Olejnik, M.~Polikarpov, S.~Syritsyn and V.~Zakharov,
  Phys.\ Rev.\ D {\bf 71} (2005) 114507
  [arXiv:hep-lat/0504008];
  J.~Greensite, A.~V.~Kovalenko, S.~Olejnik, M.~I.~Polikarpov, S.~N.~Syritsyn and V.~I.~Zakharov,
  Phys.\ Rev.\ D {\bf 74} (2006) 094507
  [arXiv:hep-lat/0606008].

\end{thebibliography}
\end{document}